\documentclass[aps,twocolumn,prc,amssymb,amsmath,bm,floatfix,superscriptaddress,nofootinbib,float]{revtex4}
\usepackage[dvipsnames,usenames]{xcolor}
\usepackage{graphicx}
\usepackage{mathptmx}
\usepackage{bm}

\newcommand{\be}{\begin{equation}}
\newcommand{\ee}{\end{equation}}
\newcommand{\bea}{\begin{eqnarray}}
\newcommand{\eea}{\end{eqnarray}}
\newcommand{\ba}{\begin{align}}
\newcommand{\ea}{\end{align}}

\newcommand{\rme}{{\rm e}}
\newcommand{\rmi}{{\rm i}}

\begin{document}

\title{Dispersion and decay of collective modes in neutron star cores}
\author{D. N. Kobyakov}
\affiliation{Institute of Applied Physics of the Russian Academy of Sciences, 603950 Nizhny Novgorod, Russia}
\affiliation{Institut f\"ur Kernphysik, Technische Universit\"at Darmstadt, 64289 Darmstadt, Germany}
\affiliation{Radiophysics Department, Lobachevsky State University, 603950 Nizhny Novgorod, Russia}
\author{C. J. Pethick}
\affiliation{The Niels Bohr International Academy, The Niels Bohr Institute, University of Copenhagen, Blegdamsvej 17,
DK-2100 Copenhagen \O, Denmark}
\affiliation{NORDITA, KTH Royal Institute of Technology and Stockholm University, Roslagstullsbacken 23, SE-106 91 Stockholm,
Sweden}
\author{S. Reddy}
\affiliation{Institute for Nuclear Theory, University of Washington, Seattle, WA 98195-1550}
\affiliation{Department of Physics, University of Washington, Seattle, WA 98195-1550}
\author{A. Schwenk}
\affiliation{Institut f\"ur Kernphysik, Technische Universit\"at Darmstadt, 64289 Darmstadt, Germany}
\affiliation{ExtreMe Matter Institute EMMI, GSI Helmholtzzentrum f\"ur Schwerionenforschung GmbH, 64291 Darmstadt, Germany}
\affiliation{Max-Planck-Institut f\"ur Kernphysik, Saupfercheckweg 1, 69117 Heidelberg, Germany}

\begin{abstract}
We calculate the frequencies of collective modes of neutrons, protons, and electrons in the outer core of neutron stars. The
neutrons and protons are treated in a hydrodynamic approximation and the electrons are regarded as collisionless.  The
coupling of the nucleons to the electrons leads to Landau damping of the collective modes and to significant dispersion of
the low-lying modes.  We investigate the sensitivity of the mode frequencies to the strength of entrainment between
neutrons and protons, which is not well characterized. The contribution of collective modes to the thermal conductivity is
evaluated.
\end{abstract}

\maketitle

\section{Introduction}

In the outer core of a neutron star, matter consists of neutrons, protons and electrons, with possibly other minority
constituents.  The protons are expected to be superconducting with pairs in the  $^1$S$_0$ state and gaps  of order 1 MeV.
The accuracy of present calculations does not permit an unambiguous answer to  the question of whether or not neutrons are
superfluid: calculations in which the internucleon interaction is taken to be the same as in free space predict pairing in
the $^3$P$_2$ state \cite{Hoffberg} while estimates that take into account the modification of the interaction by the medium
predict very small gaps \cite{SchwenkFriman} and are consistent with there being no gap (for a review of pairing gaps in
neutron star matter see \cite{Gezerlis}).  Collective modes  have previously been studied in Refs.~\cite{Epstein,BaldoDucoin2003,BedaqueReddy2014, KobyakovEtal2015}, and the possible role of such modes in transporting heat in neutron stars has been
discussed in Ref.~\cite{Aguilera2009}.
In this work, we shall assume that matter consists only of superfluid neutrons, superconducting protons and normal electrons.  The
superfluid critical temperatures for both neutrons and protons are estimated to be of the order of $1\,\mathrm{MeV}$, which
is about two orders of magnitude larger than the typical temperature in the interior of neutron stars, and therefore it is a
good first approximation to take the temperature to be zero.

In a recent work \cite{KobyakovCJP} collective modes with wavelengths large compared with the Debye screening length for
the electrons were studied under the assumption that the electrons could be treated in the hydrodynamic approximation.   Here
we extend that treatment to shorter wavelengths.  When the wavelength is of order or shorter than the Debye screening length
for electrons, the modes exhibit dispersion.  In addition, modes are damped by collisions between electrons when the
electrons are in the hydrodynamic regime and by Landau damping when the electrons are collisionless \cite{Landau1945}.
On the basis of our results for the damping of collective modes, we calculate the contribution of collective modes to the
thermal conductivity of the outer core.

The plan of this paper is as follows.  In Section II, we extend the formalism previously used to calculate the velocities of
collective modes at long wavelengths \cite{KobyakovCJP} to wavelengths shorter than the Debye screening length, when the
plasma is no longer locally neutral electrically.   Expressions for the real and imaginary parts of mode frequencies are
given in Sec.~III, and in Sec.~IV we discuss the magnitude of the parameter determining the entrainment of neutrons by
protons (and vice versa) and give numerical results for the mode frequencies derived using an equation of state based on
chiral effective field theory \cite{HebelerEtal2013}.
The thermal conductivity of matter is calculated in Sec.~V, where we also comment on possible
astrophysical effects.  In the Appendix we demonstrate that the approach adopted in this paper, where we work in terms of the
nucleon momenta per particle in the condensates, is equivalent to the traditional two-fluid formalism \cite{CJPChamelReddy,ChamelPageReddy,KobyakovPethick2013}, where one works in
terms of the momentum per particle in the neutron condensate and the velocity of the charged particles.

\section{Basic formalism}

We shall assume that the neutrons are superfluid, and that the protons are superconducting and we denote the phases of the pair
wave functions by $2\phi_n$ and $2\phi_p$, respectively.  Here we shall not take into account variations in the spatial
orientation of possible anisotropic order parameters.  We shall work in terms of the momenta per particle in the neutron and
proton condensates, which are given by
\be
{\bf p}_\alpha={\hbar}\bm\nabla \phi_\alpha,
\ee
where the index $\alpha=n,p$ refers to neutrons and protons.
In the conventional treatment of superfluid dynamics, one defines quantities ${\hbar}\bm\nabla \phi_\alpha/m$, where $m$ is
the particle mass.
These are referred to as superfluid velocities, even though they are covariant vectors, and therefore do not transform as
velocities, which are contravariant vectors.
To avoid confusion, we shall work almost exclusively in terms of the ${\bf p}_\alpha$.
We shall neglect the difference between the neutron and proton rest masses since it is small compared with other effects,
such as the contribution of interactions to the mass density, which we also neglect. Our approach is to start
from the equations of motion for the nucleon densities and condensate phases, but we shall not impose the
condition that the proton and electron densities are equal locally.  Rather we shall include the Coulomb interaction between
electrons and protons
explicitly.
For frequencies small compared with the pairing gaps and for wavelengths large compared with the coherence lengths, the
neutrons and protons may be described by hydrodynamics (see \cite{KobyakovCJP} and references therein), and the Hamiltonian for the system is thus
\bea
H&=&\int d^3r \left(E^{\rm nuc}(n_n,n_p) + \frac{n_{nn}}{2m} p_n^2 +\frac{n_{nn}}{2m}p_p^2+\frac{n_{np}}{m}\mathbf{p}_n\cdot
\mathbf{p}_p \right)\nonumber \\
&&+H^e_{\rm kin} +H^{\rm Coul}.
\eea
Here the superfluid density tensor is given by
\be
n_{\alpha \beta} =  m\frac{\partial^2E}{\partial {\mathbf p}_\alpha\partial {\mathbf p}_\beta},
\ee
where $E$ is the energy density.
Here we include terms of second order in the velocities and neglect higher order contributions.
With the help of the conditions \citep{Mendell1991a,BorumandJoyntKluzniak1996}
\bea
n_{pp}+n_{np} =n_p
\label{n_pp}
\eea
and
\bea
n_{nn}+n_{np} =n_n,
\eea
which follow from the Galilean invariance of the neutrons and protons when relativistic effects are neglected, the kinetic
energy contributions to the Hamiltonian density take on the form
\bea
\frac{n_{nn}}{2m} p_n^2 +\frac{n_{pp}}{2m}p_p^2+\frac{n_{np}}{2m}\mathbf{p}_n \cdot \mathbf{p}_p \nonumber \\
=\frac{n_n}{2m} p_n^2  +\frac{n_p}{2m}p_p^2-\frac{n_{np}}{2m}(\mathbf{p}_n - \mathbf{p}_p)^2.
\eea
When the condensation energy associated with pairing is small, the $n_{\alpha \beta}$ may be expressed in terms of Landau
parameters \cite{BorumandJoyntKluzniak1996}:
\begin{eqnarray}
n_{nn}=n_n\frac{m}{m_n^*}\left(  1+N_n(0)\frac{f_1^{nn}}{3} \right),
\label{eq:n_nn}
\end{eqnarray}
\begin{eqnarray}
n_{pp}=n_p\frac{m}{m_p^*}\left(  1+N_p(0)\frac{f_1^{pp}}{3} \right),
\label{eq:n_pp}
\end{eqnarray}
and
\begin{eqnarray}
n_{np}=\frac{m~ k_n^2~ k_p^2}{9\pi^4} f_1^{np}.
\label{eq:n_np}
\end{eqnarray}
Here $k_\alpha$ is the Fermi wavenumber, $m_\alpha^*$ the effective mass and $N_\alpha(0)=m^*_\alpha k_\alpha/\pi^2\hbar^2$ the
density of states at the Fermi surface of species $\alpha$, and $f_1^{\alpha\beta}$ is the $l=1$ component of the
quasiparticle interaction.
The equations of motion for the neutrons and protons may be obtained from Hamilton's equations, since the particle density
$n_\alpha$ and $\hbar \phi_\alpha$ are conjugate variables and therefore $\dot{n}_\alpha=  \hbar^{-1}\delta H/\delta
\phi_\alpha$ and $\dot{\phi}_\alpha=-\hbar^{-1}\delta H/\delta n_\alpha$ \cite{LL1980}.
The continuity equations for the number densities of neutrons and protons have the same form as in
Ref.~\cite{KobyakovCJP}:
\be
 \partial_t n_n+\bm{\nabla}\cdot\left[\frac{n_n}{m} \mathbf{p}_n-\frac{n _{np}}{m}\left(
\mathbf{p}_n-\mathbf{p}_p \right) \right]=0
\label{contn}
\ee
and
\be
 \partial_t n_p+\bm{\nabla}\cdot\left[ \frac{n_p}{m} \mathbf{p}_p-\frac{n _{np}}{m}\left(
\mathbf{p}_p-\mathbf{p}_n \right) \right]=0.
\label{contp}
\ee

In this paper we shall consider only linear modes.  In this case, the time dependence of the phases is given by
\be
\label{Euln}
  \hbar{{\partial }_{t}}{{\phi }_{n}}= -\mu^{\rm nuc}_n
\ee
and
\be
\label{Eulp}
 \hbar{{\partial }_{t}}{{\phi }_{p}}= -\mu^{\rm nuc}_p -e \Phi,
\ee
where
\be
\mu^{\rm nuc}_\alpha=\frac{\partial E^{\rm nuc}}{\partial {n_\alpha}}
\ee
is the nucleon chemical potential in the absence of Coulomb contributions and in the absence of flows. The quantity $\Phi$ is
the electric potential, which satisfies the
Poisson equation,
\be
\nabla^2\Phi=-4\pi e(n_p-n_e),
\ee
since the total charge density is $e(n_p-n_e)$.  Therefore the equations of motion for ${\mathbf p}_\alpha$ are
\be
\label{Eulp1}
\partial_t{\mathbf p}_p
= -\bm{\nabla}\left(\mu^{\rm nuc}_p +e \Phi  \right)
\ee
and
\be
\label{Euln1}
{{\partial }_{t}}{{\mathbf p}_{n}}= -\bm{\nabla}\mu^{\rm nuc}_n.
\ee

To close this set of equations one needs to calculate the response of the electrons to an electric potential.  We shall
consider small disturbances from the uniform state and therefore one can use linear response theory.  If the frequency,
$\omega$, of the  disturbance is large compared with the electron collision frequency and the wave number is large compared
with the inverse mean free path, the electrons may be treated as collisionless.  The mean free path of
electrons is limited  by scattering from electrons, and for calculating the properties of longitudinal modes, the relevant
relaxation time is that for viscosity.   This has been calculated in detail by Shternin and Yakovlev
\cite{ShterninYakovlev2008}, who demonstrate that the dominant scattering process for electrons at temperatures less than
$\sim ~ 10^9$ K is via exchange of transverse photons, rather than the Coulomb interaction.  For nuclear matter density and a
temperature of $10^8$K, their calculations lead to a mean free path for viscosity of $\sim 10^{10}$ fm or $10^{-3}$ cm.
This is intermediate between typical stellar length scales ($\sim$ km) and the wavelength scale of thermal excitations $\sim
\hbar v/(k_B T) \approx 10^4(v/c)/T_8$ fm.     Here $v$ is the velocity of the mode, $k_B$ is the Boltzmann constant, and
$T_8$ is the temperature in units of $10^8$ K.  In this paper, our primary interest is collective modes on a microscopic
scale, so we shall assume that wavelengths are short compared with the electron mean free path and frequencies are large
compared with the electron collision frequency and we shall henceforth neglect collisions.

In the absence of collisions, the response of electrons in dense matter is given to a very good approximation by the random
phase approximation and  for ultrarelativistic electrons the result is \cite{Jancovici1962}
\be
\delta n_e=\chi_0  e \delta \Phi.
\ee
Here
\bea
\chi_0&=&\frac{\partial n_e}{\partial \mu_e}\left(1-\frac{s}{2}\ln\frac{s+1}{s-1}\right) \nonumber \\
&=&\frac{\partial n_e}{\partial \mu_e} \left(1-\frac{s}{2}\ln\left|\frac{s+1}{s-1}\right|+\rmi
\frac{\pi}{2}s~\Theta(1-|s|)\right),
\eea
with $s=\omega/(ck)$, $\partial n_e/\partial \mu_e =p_e^2/\pi^2\hbar^3c$ and $\Theta$ is the Heaviside step function.
This form is valid for $\omega\ll \mu_e$ and $k\ll k_e$.  In the absence of an external potential, the electric potential is
given by
\be
\Phi=\frac{4\pi e}{k^2}(n_p-n_e),
\label{Phi}
\ee
and therefore to linear order
\be
\delta n_e= \frac{4\pi e^2}{k^2}\chi_0(\delta n_p-\delta n_e)
\ee
or
\be
\delta n_e =\frac{(4\pi e^2/k^2)\chi_0}{1+(4\pi e^2/k^2)\chi_0}\delta n_p=\frac{(4\pi
e^2/k^2)\chi_0}{\epsilon_e(\omega,k)}\delta n_p.
\label{electrondens}
\ee
Here
\be
\epsilon_e(\omega,k)=1+(4\pi e^2/k^2)\chi_0
\label{epsilone}
\ee
 is the dielectric function of the electrons.
Equation (\ref{electrondens}) enables us to eliminate $\delta n_e$ from the equations of motion.

\section{Collective modes}

The properties of collective modes are obtained by solving Eqs.\ (\ref{contn}) -- (\ref{Eulp}) together with Eqs. (\ref{Phi})
and (\ref{electrondens}).  Small longitudinal perturbations about a uniform state in which the nucleons and electrons are
stationary satisfy the equation
\begin{equation}
\label{matrix_formula}
\left(
                     \begin{array}{cccc}
                       -{m\omega}/{k} & 0 & n_{pp} & n_{np} \\
                       0 & -{m\omega}/{k} & n_{np} & n_{nn} \\
                       E_{pp} & E_{np} & -{\omega}/{k} & 0 \\
                       E_{np} & E_{nn} & 0 & -{\omega}/{k}
                     \end{array}
                   \right)\left(\begin{array}{c}
  \delta n_p \\
  \delta n_n \\
  \delta p_p \\
  \delta p_n
\end{array}\right)=0,
\end{equation}
where
\be
E_{pp}= E_{pp}^{\rm nuc} +\frac{4\pi e^2}{k^2 \epsilon_e(\omega,k)}.
\label{E_pp}
\ee
Here, $E_{pp}^{\rm nuc}=\partial\mu_p^{\rm nuc}/\partial n_p$ and $E_{n\alpha}=\partial\mu_n^{\rm nuc}/\partial n_\alpha$.
The first term on the right side of Eq.~(\ref{E_pp}) represents the contribution from nuclear interactions and the second
term is the Coulomb interaction between protons screened by electrons.
We have dropped the superscript ``nuc'' on the other $E_{\alpha\beta}$ because there are no Coulomb contributions.
Equation (\ref{matrix_formula}) reduces to the result of Ref.\  \cite{KobyakovCJP} if one neglects the frequency
dependence of the electron response, and replaces $\epsilon_e$ by its asymptotic long-wavelength form $\simeq k_{TF}^2/k^2$,
where $k_{TF}=\sqrt{4 \pi e^2\partial n_e/\partial \mu_e}$ is the Thomas--Fermi screening wave number.

The dispersion relation for the collective modes is given by the condition $\det M=0$, where $M$ is the matrix in Eq.\
(\ref{matrix_formula}).  In general, there are three collective modes, corresponding to the three components:  two acoustic
modes with in-phase and out-of-phase motions of the neutrons and the charged particles, and the plasma oscillation of the
electrons, which has a nonzero frequency $\approx (4\pi n_e e^2c^2/\mu_e)^{1/2}$ as $k\to 0$.   The present formalism is
inadequate to calculate the frequency shift of the electron plasma oscillation due to the motion of neutrons and protons
because the frequency is high compared with the gap frequencies and consequently the hydrodynamic approximation for the
neutrons and protons is not valid.

\subsection{Uncoupled modes}
It is instructive to investigate the case when there is no coupling between neutrons and protons ($E_{np}=0, n_{np}=0,
n_{pp}=n_p,$ and $n_{nn}=n_n$).   The dispersion relation for the collective mode of electrons and protons is then
\begin{equation}
  \omega^2=\frac{\Omega_p^2}{\varepsilon_e(\omega,k)}+\frac{n_pE_{pp}^{\rm nuc}}{m}k^2\equiv v_{p0}^2k^2,
\label{plasmon}
\end{equation}
 where $\Omega_p=({4\pi e^2 n_p}/{m })^{1/2}$  is the proton plasma frequency.

Because of the frequency- and wave number dependence of $\epsilon_e$, Eq.~(\ref{plasmon}) has  two solutions, the electron
plasma oscillation and an acoustic mode, which has a velocity small compared with $c$.
To calculate the frequency of the latter mode we may expand $\chi_0$ in powers of $s$,
\be
\chi_0\simeq\frac{\partial n_e}{\partial \mu_e} \left(1+\rmi\frac{\pi}{2}s-s^2\right).
\label{chi0}
\ee
As we shall show, to take into account the nonzero effective mass of the electrons it is necessary to retain terms of
order $s^2$.
We write the frequency of the mode in terms of its real part $\omega'$ and its imaginary part $\omega''$.   In the long-wavelength limit, one finds
\begin{eqnarray}
\left[m+\frac{\mu_e}{3c^2}\left(\frac{\pi^2}{4}-1\right)\right]\omega'^2 \simeq
\frac{\mu_e}{3}k^2+n_pE_{pp}^{\mathrm{nucl}}k^2.
\label{omegalf1}
\end{eqnarray}
The coefficient of $\omega^2$ in Eq.~(\ref{omegalf1}) shows that an electron contributes an amount
$(\pi^2/4-1)\mu_e/3c^2\approx 0.49\mu_e/c^2$ to the effective mass of the charged particles.

If one further neglects $E^{\rm nuc}_{pp}$ and considers the long-wavelength limit $k\to0$. Eq.~(\ref{omegalf1})
is the analog for relativistic electrons of the expression for the sound speed in metals derived by Bohm and Staver \cite{BohmStaver}.
The leading contribution to $\omega''$ is
\be
\omega'' \simeq   -\frac{\pi}{12}\frac{\mu_e}{mc^2}ck.
\label{omegalfim}
\ee
This is due to Landau damping, the decay of the mode into
electron--hole pairs \cite{Landau1945}.

If collisions between electrons are so frequent that conditions are hydrodynamic, the response function for the electrons has
the form
\be
\chi=\frac{\partial n_e}{\partial \mu_e}\frac{c_e^2k^2}{c_e^2k^2-\omega^2},
\ee
where $c_e=c/\sqrt3$ is the speed of hydrodynamic (first) sound in an ultrarelativistic Fermi gas.
With this expression for $\chi$, one sees that the effective mass of an electron is $\mu_e/c^2$, or roughly twice the value
for the collisionless limit.
For $E_{pp}^{\rm nuc}\neq 0$, $\mu_e$ in Eqs. (\ref{omegalf1})-(\ref{omegalfim}) is replaced by $\mu_e+3E_{pp}^{\rm
nuc}/n_p$.  Since $\mu_e\sim$ 100 MeV, the effects of the electron inertia in neutron star cores are at the 10\% level  and
are therefore much more important than in terrestrial matter.
However, the effects are comparable to the deviations of the energy per particle from the rest mass, which we have neglected
in this article, where we have treated the nucleons nonrelativistically.

Due to incomplete screening at higher wave numbers, the velocity of the proton--electron mode depends on $k$.
On solving the eigenvalue problem to first order in $\epsilon_e''=\Im \epsilon_e$ one finds
\be\label{omegaepsil1expl}
\omega_0'^2=\frac{n_{p}}{m}\left(E_{pp}^{\rm nuc}+\frac{4\pi e^2}{k^2+k_{TF}^2}\right)k^2,\\
\ee
and
\be
\label{omegaepsil2expl}
\omega_0''=-\frac{\pi}{12}\frac{\mu_e}{mc^2}\frac{ck}{(1+k^2/k_{TF}^2)^2},
\ee
which shows that the velocity and damping of modes is markedly reduced for $k\gtrsim k_{TF}$.
Equation (\ref{omegaepsil1expl}) shows that both nucleons and electrons contribute to the bulk modulus of the medium.

The frequency of the neutron mode when coupling between neutrons and protons is neglected is given by
\be
\omega^2=\frac{n_nE_{nn}}{m}k^2\equiv v_{n0}^2k^2.
\label{vn0}
\ee

\subsection{The general case}
The solution of  the eigenvalue problem for Eq. (\ref{matrix_formula})  written in terms of the phase velocities $v=\omega/k$
of the modes has the form
\begin{equation}
\label{omega_coup}
v_\pm^2=\frac{c_s^2}{2} \pm\sqrt{\left(\frac{c_s^2}{2}\right)^2-\frac{\det[n_{\alpha \beta}]\det[E_{\alpha \beta}]}{m^2}},
\end{equation}
where
\begin{align}\label{c_s}
c_s^2=(E_{pp}n_{pp}+2E_{np}n_{np}+E_{nn}n_{nn})/m \nonumber \\
=v_{p0}^2+v_{n0}^2+\frac{n_{np}}{m}(2E_{np} -E_{nn}-E_{pp}).
\end{align}
where $v_{p0}$ and $v_{n0}$ are the velocities of the modes in the absence of coupling between neutrons and protons, given by
Eqs. (\ref{plasmon}) and (\ref{vn0}).
To bring out the effects of the entrainment, it is useful to calculate mode frequencies in the absence of entrainment,
\begin{equation}\label{vsc}
  {\left(v_\pm^{sc}\right)}^2=\frac{v_{p0}^2+v_{n0}^2}{2}\pm\sqrt{\left(\frac{v_{p0}^2-v_{n0}^2}{2}\right)^2+\frac{n_nn_pE_{np}^2}{m^2}}.
\end{equation}
The supercript $sc$ indicates that only the scalar coupling between components is present, and the vector coupling that gives rise to entrainment is absent.
Equation (\ref{omega_coup}) is algebraically equivalent to the results derived for collective mode velocities in the inner crust of a neutron star in
Ref.\ \cite[Sec. III]{KobyakovPethick2013} from the standard two-fluid model.  The fact that, in the neutron star crust, the
charged particles are not superfluid is irrelevant for the mode frquencies, because in that work the ions and electrons were treated
using a hydrodynamic approtach.
From Eq.~(\ref{omega_coup}) one sees that there is a zero frequency mode if $\det[E_{\alpha \beta}]$ or $\det[n_{\alpha
\beta}]$ vanish.  This reflects the fact that the uniform system with no flow of protons or neutrons is unstable to formation
of a density wave if $\det[E_{\alpha \beta}]=0$ and to relative flow of the neutrons and protons if $\det[n_{\alpha
\beta}]=0$.

In the long-wavelength limit the imaginary part of $E_{pp}$ can be obtained by combining Eqs. (\ref{epsilone}), (\ref{E_pp}) and (\ref{chi0}) and Taylor-expanding $\chi_0^{-1}$: $E_{pp}''=E_{ee}''\approx-(\pi/2) E_{ee}'\omega/ck$, where the real part of $E_{ee}$ is $E_{ee}'=\partial\mu_e/\partial n_e$.
Thus, the dispersion relation for the modes, the condition that the determinant of the matrix in Eq.\ (\ref{matrix_formula}) vanish, may be written
\be
(v^2-{\tilde v}_+^2)(v^2-{\tilde v}_-^2)  -\frac{\pi E_{ee}'}{2} \left(E_{nn}\frac{\det n_{\alpha\beta}}{m^2} -\frac{n_{pp}}{m}v^2 \right)\frac{v}{c}=0,
\ee
where ${\tilde v}_\pm$ are the mode velocities when $E_{ee}''=0$. To first order in $E_{ee}''$ the real parts of the velocities are equal to $v_\pm'$ and the imaginary parts of the velocities are thus given by
\be
v_+''=-\frac{\pi E_{ee}'}{2c}\frac{\left[(n_{pp}/m)v_+^2 -E_{nn}{\det n_{\alpha\beta}}/{m^2} \right]}{(v_+^2-v_-^2)}
\label{v''+}
\ee
and
\be
v_-''=-\frac{\pi E_{ee}'}{2c}\frac{\left[ E_{nn}\det n_{\alpha\beta}/{m^2}-(n_{pp}/m)v_-^2 \right]}     {(v_+^2-v_-^2)},
\label{v''-}
\ee
where the $v_\pm$ on the right sides are to be evaluated in the absence of damping.

\subsection{Physical character of the modes}
The coupled modes generally involve oscillations of both the neutron and proton densities.
On eliminating the momentum variables $p_p$ and $p_n$ from the linear system given in Eq. (\ref{matrix_formula}), we find

\begin{eqnarray}
\nonumber
\left(
                     \begin{array}{cc}
                       -{mv^2}+n_{nn}E_{nn}+n_{np}E_{np} & n_{nn}E_{np}+n_{np}E_{pp} \\
                       n_{np}E_{nn}+n_{pp}E_{np} & -{mv^2}+n_{np}E_{np}+n_{pp}E_{pp}
                     \end{array}
                   \right)\\
\label{matrix_reduced}
                   \times\left(\begin{array}{c}
  \delta n_n \\
  \delta n_p
\end{array}\right)=0.\quad
\end{eqnarray}
Therefore the ratio of the neutron and proton density variations in the eigenmode with velocity $v_\pm$ is given by
\begin{equation}\label{ratio_ampl}
  \left.\frac{\delta n_n}{\delta n_p}\right|_{v_\pm}=\frac{n_{nn}E_{np}+n_{np}E_{pp}}{{mv_\pm^2}-n_{nn}E_{nn}-n_{np}E_{np}}.
\end{equation}
It is convenient to define mixing angles by the equations
\begin{equation}\label{thetaplus}
  \gamma_{n+}=\arctan{\left.\frac{\delta n_n}{\delta n_p}\right|_{v_+}}
\end{equation}
and
\begin{equation}\label{thetaminus}
  \gamma_{p-}=\arctan{\left.\frac{\delta n_p}{\delta n_n}\right|_{v_-}}.
\end{equation}
If $\gamma_{n+}=0$, the mode $v_+$ involves motion only of the charged particles.
When $0<\gamma_{n+}<\pi/2$, the mode $v_+$ corresponds to an in-phase oscillation of the proton and neutron densities, and when
$-\pi/2<\gamma_{n+}<0$, to an out-of-phase oscillation.
Analogously, the mode with velocity $v_{-}$ is a pure neutron oscillation when $\gamma_{p-}=0$, an in-phase oscillation of the neutron and proton
densities when $0<\gamma_{p-}<\pi/2$, and an out-of-phase oscillation when $-\pi/2<\gamma_{p-}<0$.

\section{Calculation of mode frequencies}
In this section we apply the formalism to calculate mode frequencies in the outer core.  For the nucleon contributions to the
quantities $E_{\alpha\beta}$ we use the results of Hebeler et al. \cite{HebelerEtal2013}, which were earlier employed in
calculations of mode frequencies at long wavelengths \cite{KobyakovCJP}.
In our numerical calculations we use values of $E_{\alpha\beta}^{nuc}$ based on chiral effective field theory, because we regard these as the best available.
However, since $f_1^{np}$ has not been calculated in detail for these interactions, we used for that the SLy4 value, Eq. (\ref{n_npSkyrme}), and estimates (\ref{n_np_m*}) and (\ref{f1}) based on arguments from nucleon effective masses obtained in chiral EFT models.
The other important piece of input is the quantity $n_{np}$ that determines the strength of entrainment.  Chamel and Haensel showed that values of $n_{np}$ obtained from Skyrme interactions had a large spread  \cite{ChamelHaensel}, and in Ref.~\cite{KobyakovCJP} the result for the SLy4 Skyrme
interaction,
\be
n_{np}^{(a)}= -1.567 \,{\rm fm}^3 n_n n_p=  -0.04012       \frac{n_n n_p}{n_0^2}     \,{\rm fm}^{-3},
\label{n_npSkyrme}
\ee
which is representative of results found for many other Skyrme interactions, was adopted.

An alternative approach to estimating $n_{np}$ is to use information on nucleon effective masses, which are linked to the
Landau parameters by the relations
\cite{Sjoeberg}
\be
\frac{m^*_n}{m} = 1 + \frac{1}{3} \frac{m^*_n k_n}{\pi^2}
\biggl[ f_1^{nn} + \biggl( \frac{k_p}{k_n} \biggr)^2
f_1^{np} \biggr]
 \label{mstarn}
\ee
and
\be
\frac{m^*_p}{m} = 1 + \frac{1}{3} \frac{m^*_p k_p}{\pi^2}
\biggl[ f_1^{pp} + \biggl( \frac{k_n}{k_p} \biggr)^2
f_1^{np} \biggr] \,. \label{mstarp}
\ee
Thus, for symmetric nuclear matter at the saturation density, it follows from  Eq.~(\ref{mstarn}) that
\bea
f_1^{np}=\frac{3\pi^2\hbar^2}{m k_0}\left(1-\frac{m}{m_0^*}\right)  -f_1^{nn},
\label{f1np_sat}
\eea
where $k_0=(3\pi^2n_0/2)^{1/3}$ is the Fermi wave number  and $m^*_0$ the nucleon effective mass for matter under those
conditions.  To obtain information about $f_1^{nn}$, we make use of results for the limiting cases of low neutron densities
and pure neutron matter.  For low neutron densities, $f_1^{nn}$ tends to zero.
Calculations for pure neutron matter indicate that $m^*_n/m \approx 1.0-1.2,~ 0.9-1.1$, and $0.8-1.0$ for densities $n=0.5
n_0, n_0$, and $1.5 n_0$,
respectively~\cite{RGnmatt,Kainmatt}. These results are all consistent with an effective mass equal to the bare mass, which
would imply that $f_1^{nn}$ for pure neutron matter is zero.  Therefore the simplest assumption for the dependence of
$f_1^{nn}$ on proton fraction is that it vanishes for all neutron densities, in which case  Eq.~(\ref{f1np_sat}) becomes
\be
f_1^{np}=\frac{3\pi^2\hbar^2}{m k_0}\left(1-\frac{m}{m_0^*}\right),
\ee
and, from Eq.\,(\ref{eq:n_np}),
\be
n_{np}=\left(1-\frac{m}{m_0^*}\right) \frac{n_0}2\,\,\,{\rm (symmetric~nuclear~matter~at~saturation)}.
\ee
 If one assumes that $f_1^{np}$ is proportional to the product of the neutron and proton Fermi wavenumbers, as it is for the
 Skyrme interaction, one thus finds
\be
n_{np}=2\left(1-\frac{m}{m_0^*}\right) \frac{n_nn_p}{n_0}.
\ee
For symmetric nuclear matter, the nucleon effective mass is typically taken to be 0.7$m$ \cite{RGnmatt, Kainmatt}, but  it could be
larger. If one is conservative and takes the value to be $0.8 m$, one finds
 \be
n_{np}^{(b)}=-\frac12 \frac{n_nn_p}{n_0}= - 3.125 ~{\rm fm}^3 {n_nn_p}=-0.0800 \frac{n_nn_p}{n_0^2}  ~{\rm fm}^{-3},
\label{n_np_m*}
\ee
roughly twice the value (\ref{n_npSkyrme}) obtained from the SLy4 Skyrme interaction.

We now estimate the changes in this result due to a nonzero value of $f_1^{nn}$.  Let us denote the effective mass of neutrons
in pure neutron matter at density $n_0$ by $m^*_{\rm NM}$.  It then follows from Eq.~(\ref{mstarn}) that
\be
(f_1^{nn})_{\rm NM}=  \frac{3\pi^2\hbar^2}{m k_{\rm NM}} \left(1-\frac{m}{m_{\rm NM}^*}\right),
\ee
where $k_{\rm NM}=(3\pi^2n_0)^{1/3}$ is the Fermi wave number of neutron matter.
If $f_1^{nn}$ scales with neutron density in the same way as it does for a Skyrme interaction ($\propto k_n^2$), this gives
for this quantity for symmetric nuclear matter at density $n_0$,
\be
f_1^{nn}=  \frac12 \frac{3\pi^2\hbar^2}{m k_0} \left(1-\frac{m}{m_{\rm NM}^*}\right).
\ee
Therefore, from Eq.~(\ref{f1np_sat}), one finds
\be
f_1^{np}=\frac{3\pi^2\hbar^2}{m k_0}\left(1-\frac{m}{m_0^*}  -\frac12  \left[1-\frac{m}{m_{\rm NM}^*}\right]\right).
\label{f1}
\ee
This shows that for $m_0=0.8\, m$, a neutron effective mass of $1.1 m$ rather than $m$ for pure neutron matter would increase
the magnitude of $f_1^{np}$, and hence also that of $n_{np}$, by $\sim 20 \%$.

Throughout these arguments, we have assumed that the Landau parameters scale with Fermi momentum as they would for a Skyrme
interaction.   How good this assumption should be investigated using microscopic calculations.  Results for low-density Fermi gases
exhibit other scalings with density, e.g., for a spin-1/2 one component Fermi gas, $f_1$ scales as the Fermi wave number, not
its square.  This is due to the fact that this term comes from the induced interaction.  However, because scattering lengths
for nucleon--nucleon scattering are so large, we do not expect the scaling behavior predicted by the low-density gas to be
relevant at most densities of interest in neutron stars.

Using the effective masses discussed above, we have evaluated the entrainment parameters defined in Eqs.~(\ref{eq:n_nn}), (\ref{eq:n_pp}), and
(\ref{eq:n_np}) and these are shown in Table \ref{tab:Eijandnij} for characteristic densities in the neutron star outer
core.  The table also shows the derivatives of the energy density $E_{\alpha\beta}$ obtained using the expression for the energy density described in
Ref.\,\cite{HebelerEtal2013}.
\begin{widetext}
\begin{center}
\begingroup
\begin{table}[htbp]
   \centering
   \resizebox{\textwidth}{!}{\begin{tabular}{cccccccccc} 
      \hline
      \hline
      $n_B ({\rm fm}^{-3})$ &  $n_{p} ({\rm fm}^{-3})$ & $n_{n} ({\rm fm}^{-3})$  & ${n_{np}}^{(a)} ({\rm fm}^{-3})$ & ${n_{np}}^{(b)} ({\rm fm}^{-3})$ & $E_{pp}^{\rm nuc}$ (GeV fm$^{3}$) & $E_{nn}$ (GeV fm$^{3}$) &  $E_{np}$ (GeV fm$^{3}$)\\
      \hline
          0.08 &  $2.64\times10^{-2}$ & $7.736\times10^{-2}$ & $-3.205\times10^{-4}$ & $-6.391\times10^{-4}$ & $1.294$ & 0.1886 & -0.5729 \\
          0.16 &  $7.746\times10^{-3}$ & $1.523\times10^{-1}$ & $-1.851\times10^{-3}$ & $-3.692\times10^{-3}$ & 1.247 & 0.2684 & -0.1891 \\
          0.24 &  $1.236\times10^{-2}$ & $2.277\times10^{-1}$ & $-4.418\times10^{-3}$ & $-8.811\times10^{-3}$ & 1.325 & 0.3352 & 0.0811 \\
          0.32 &  $1.513\times10^{-2}$ & $3.048\times10^{-1}$ & $-7.239\times10^{-3}$ & $-1.444\times10^{-2}$ & 1.433 & 0.3895 & 0.2972 \\
      \hline

      \end{tabular}}

     \caption{Beta-equilibrium densities, entrainment parameter and thermodynamic derivatives. The derivatives $E_{\alpha\beta}$ are calculated from an equation of state based on chiral effective field theory \cite{HebelerEtal2013}.
     The entrainment parameter ${n_{np}}^{(a)}$ is calculated from Fermi liquid theory and Skyrme energy functional SLy4, Eq.\,(\ref{n_npSkyrme}), while  ${n_{np}}^{(b)}$ is derived from considerations of nucleon effective masses, Eq.\,(\ref{n_np_m*}).}
   \label{tab:Eijandnij}
\end{table}
\endgroup
\end{center}
\end{widetext}

We write the mode velocity in terms of its real part, $v'$, and its imaginary part $v''$ and results for these quantities are  shown in Figs. 1 and 2, respectively.
Only Landau damping is included in calculating the imaginary
parts. This is valid when $ k <  2 \Delta/v' $, since for larger wave numbers pair-breaking is possible.

Equation (\ref{omega_coup}) determines the real and imaginary parts of the eigenmode velocities and from them one may obtain the mean free paths,
\be
l_k=v' \tau_k=\frac12\frac{v'}{v''}\frac1k=    \frac1{4\pi}\frac{v'}{v''}\lambda=  \frac12\frac{(v')^2}{v''}\frac1{\omega_k},
\label{mfp}
\ee
where $\tau_k=1/{2kv''}$ is the relaxation time of the hydrodynamic motion. The factor of $1/2$ comes from the fact that the relaxation of the hydrodynamic motion is linked to damping of the \emph{intensity} of the mode, which is the amplitude squared, while the amplitude is exponentially damped on a time scale $1/{kv''}$.
At nuclear matter density and at long wavelengths for a typical thermal phonon frequency $\omega=3k_BT/\hbar$ with $T=10^8$ K, the mean free path of the predominantly neutron mode is $3.022\times10^{-9}$ cm for the case of no entrainment, $3.248\times10^{-9}$ cm for $n^{(a)}_{np}$ and $3.321\times10^{-9}$ cm for $n^{(b)}_{np}$, while for the predominantly charged particle mode the corresponding results are $7.003\times10^{-10}$ cm, $6.454\times10^{-10}$ cm, and $6.492\times10^{-10}$ cm.
The mean free paths for a thermal excitation vary as $1/T$ and numerically they are much shorter than
those due to phonon--phonon scattering, which diverge as a higher inverse power of $T$ at low temperatures
\cite{ManuelTolos2011}.

For weak mixing, $E_{np}^2 \ll {E_{nn} E_{pp}}$ and $n_{np}^2 \ll {n_n n_p}$,  we can obtain from Eqs.~(\ref{v''+}) and (\ref{v''-}) simple expressions for the damping of the modes.  
Here and in the remainder of this section $E_{pp}$ and $E_{ee}$ refer to the real parts of these quantities.
For the predominantly charged-particle mode one finds
\be
v_p''  \simeq -\frac{\pi}{4mc}n_pE_{ee}
\ee
and the mean free path is given by
\bea
&&l_p(\omega)=\frac{2}{\pi}\frac{E_{pp}}{E_{ee}} \frac{c}{\omega}\nonumber \\
&&\simeq 1.66 \times 10^{-8} \left(\frac{\mu_e}{100~{\rm MeV}} \right)^2 \left(\frac{E_{pp}}{10^4~{\rm MeV~fm^3}}\right)
\left(\frac{\rm keV}{\hbar \omega} \right) ~{\rm cm}\,.\nonumber \\
\eea
For the predominantly neutron mode the corresponding results are
\be
v_n''  \simeq -\frac{\pi \zeta^2}{4mc}n_nE_{ee}    = \zeta^2\frac{n_n}{n_p} v_p'' ,
\ee
where
\be
\zeta=\frac{ n_{p} E_{pn}+n_{np} E_{nn}}{n_{p} E_{pp}-n_{n} E_{nn}},
\ee
and
\bea
&& l_n(\omega)=\frac{2}{\pi \zeta^2}\frac{E_{nn}}{E_{ee}}\frac{c}{\omega} \nonumber \\
  &&\simeq \frac{1.66 \times 10^{-8}}{\zeta^2} \left(\frac{\mu_e}{100~{\rm MeV}} \right)^2 \left(\frac{E_{nn}}{10^4~{\rm
  MeV~fm^3}}\right)  \left(\frac{\rm keV}{\hbar \omega} \right) ~{\rm cm}.\nonumber \\
\eea
The results have limited utility for the situation considered in this paper because there is an avoided crossing of the two sound speeds, as one can see from Fig.~3.  In the  vicinity of the avoided crossing, the mixing is strong.  Another complicating feature is that the quantity  $ n_{p} E_{pn}+n_{np} E_{nn}$ that enters in the expression for $\zeta$  passes through zero at a density close to that of the avoided crossing.  This results in vanishing of the damping of the lower frequency mode (and a divergence of the mean free path) at a particular density, as may be seen in Fig.\,4.

In earlier work,  Bedaque and Reddy \cite{BedaqueReddy2014}, obtained a rough estimate of these mean free paths by
neglecting effects due to entrainment. The range of values they predicted is compatible with the detailed estimates presented here. It is also worth noting that in the neutron star inner crust, where phonons of the neutron superfluid mix
with the longitudinal lattice phonons, the predominantly neutron mode with thermal energy was found to have a mean free path
in the range $10^{-6}$--$10^{-3}$ cm,  much larger than that associated with lattice phonon modes, suggesting that  heat
transport due to phonons could play a role \cite{Aguilera2009}. However, revised estimates which included the effects due to
entrainment lead to larger mixing and stronger damping of the neutron mode \cite{ChamelPageReddy}.

\begin{figure}
\includegraphics[width=3.5in]{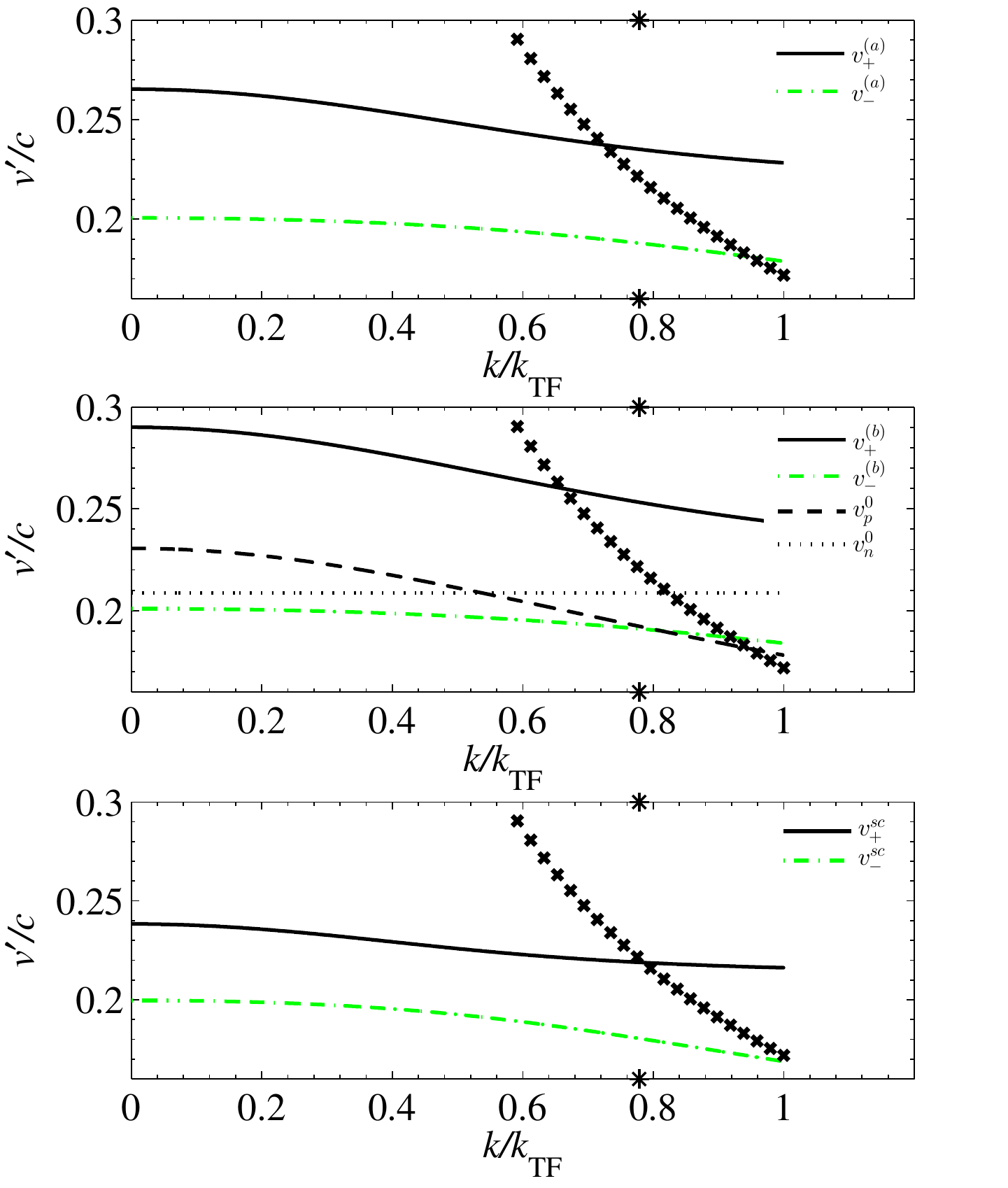}
\caption{(Color online) The real part of the mode velocity  $v$   in the collisionless  regime at nuclear saturation density $0.16~\mathrm{fm}^{-3}$, as a function of wave number.
The velocities $v_\pm^{(a,b)}$, Eq. (\ref{omega_coup}), are calculated with $n_{np}^{(a)}$, Eq. (\ref{n_npSkyrme}), or $n_{np}^{(b)}$, Eq. (\ref{n_np_m*}).
The velocities $v_{\pm}^{sc}$ are given in Eq. (\ref{vsc}), and the velocities $v_{n}^{0}$ and
$v_{p}^{0}$ of the uncoupled modes are given in Eqs. (\ref{plasmon}) and (\ref{vn0}).
The crosses correspond to the onset of damping by breaking of Cooper pairs.  This occurs for $\omega/k\geq2\Delta/\hbar k$, and the crosses correspond to a neutron or proton gap
$\Delta=1$ MeV.
The stars mark the wave number corresponding to the inverse size of the Cooper pairs of neutrons in the BCS approximation,
$\xi_n^{-1}=\pi m \Delta_n/\hbar k_n$, where the neutron gap $\Delta_n$ is taken to be 1 MeV; the inverse
size of the proton Cooper pairs lies outside the domain of the figure, $\xi_p^{-1}>k_{\mathrm{TF}}$.}
\label{fig:v'}
\end{figure}

\begin{figure}
\includegraphics[width=3.5in]{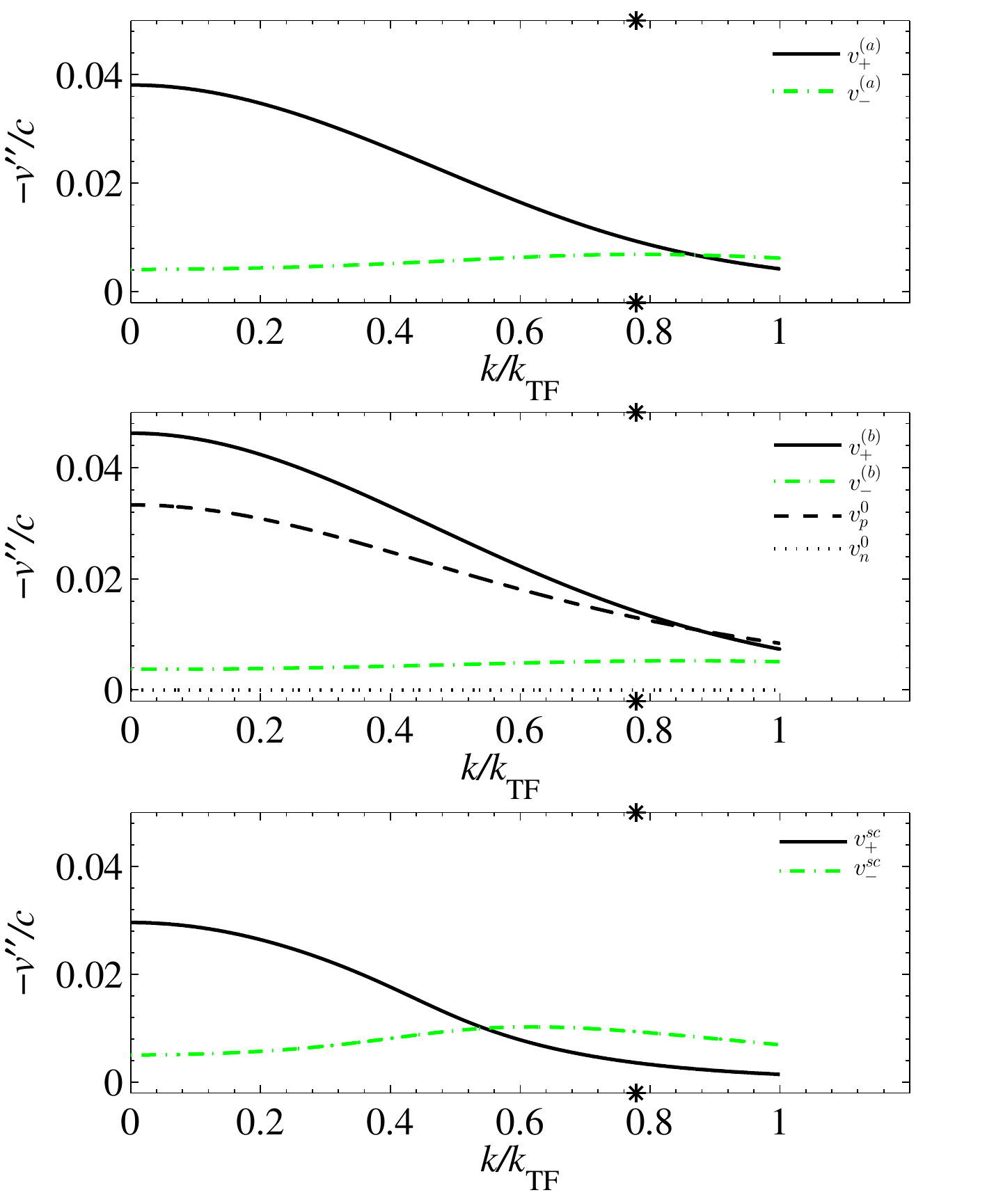}
\caption{(Color online) The imaginary part of the mode velocity  $v$  as function of wave number, at nuclear saturation density.
The modes are labeled as in Fig.\,1.
The stars mark the wave number corresponding to the inverse size of the Cooper pairs of neutrons in BCS approximation with the
neutron gap $1$ MeV.}
\label{fig:v'}
\end{figure}

\begin{figure}
\includegraphics[width=3.5in]{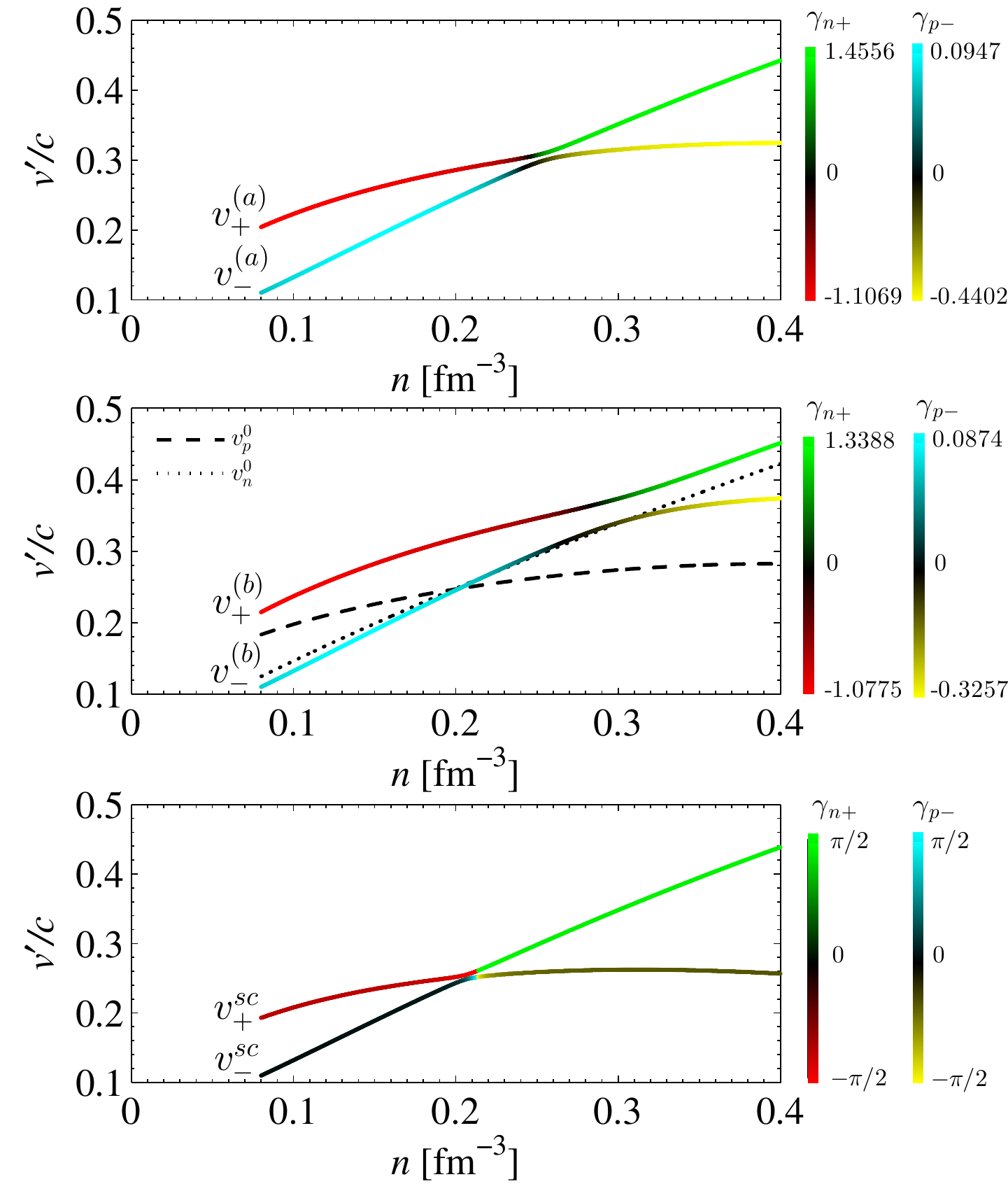}
\caption{(Color online) The real part of the collective mode velocities as function of baryon density, at long wavelengths in the collisionless limit.
The modes are labeled as in Fig. 1.
The low-density boundaries of the plots correspond to the density of the crust--core boundary, $\approx 0.08$ fm$^{-3}$
\cite{HebelerEtal2013}.
The color indicates the value of the mixing angles defined in Eqs. (\ref{thetaplus}) and (\ref{thetaminus}).}
\end{figure}

\begin{figure}
\includegraphics[width=3.5in]{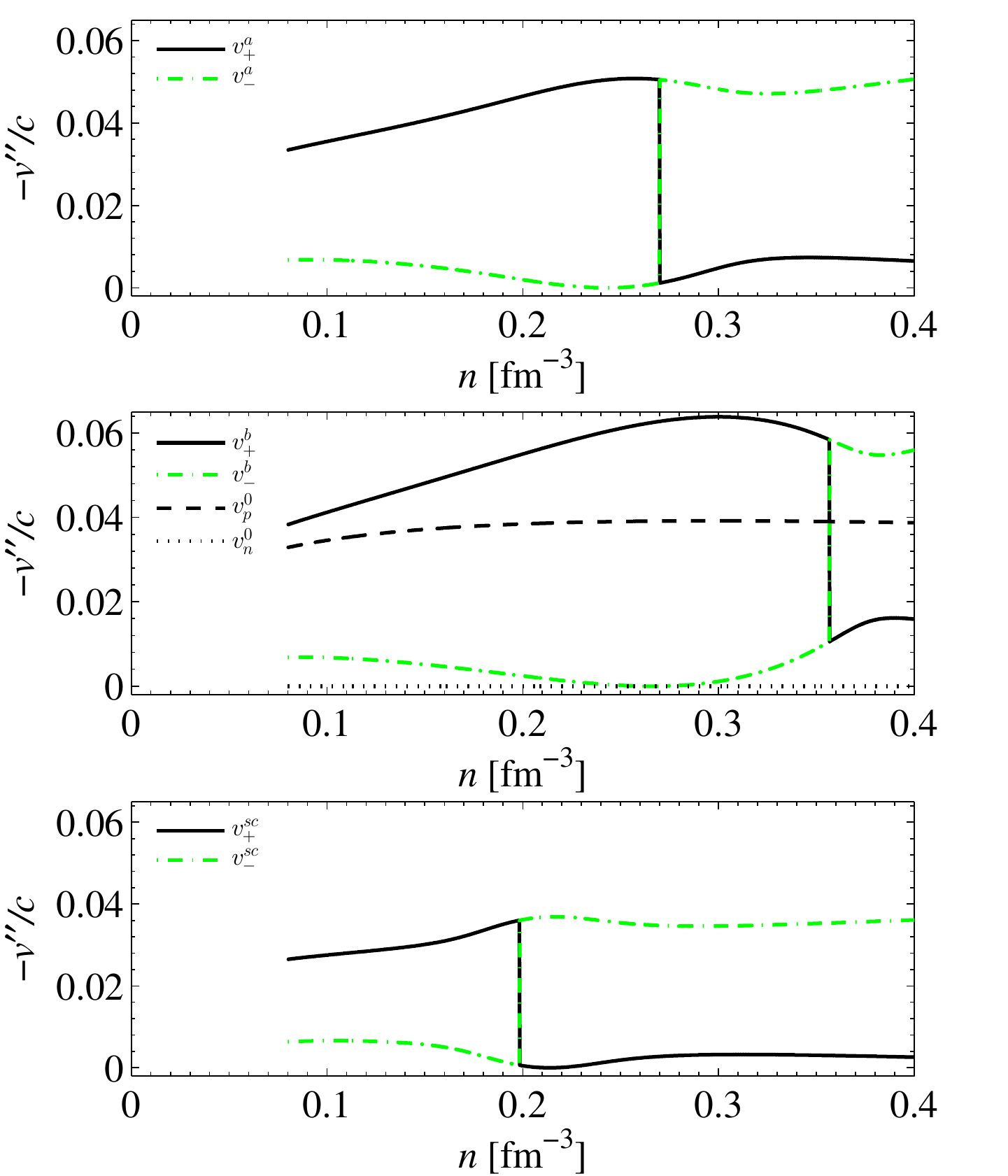}
\caption{(Color online) The imaginary part of the collective mode velocities as functions of baryon density.
The modes are labeled as in Fig. 1.}
\end{figure}

\section{Astrophysical considerations and concluding remarks}
\label{Conclusion}

The dispersion and decay of collective modes in superfluid and superconducting dense matter can influence observable aspects
of neutron star evolution. At long wavelengths, these modes are relevant to understanding neutron star seismology. Short
wavelength modes can contribute to the  heat capacity and thermal conductivity, both of which play a role in the thermal
evolution. In the following we shall explore some of the implications of the results obtained in the preceding section for
neutron star phenomenology.

The temperatures of interest in isolated neutron stars (older than a few hundred years) are expected to be in the range
$10^7$-$10^9$ K. In the core where nucleons condense to form Cooper pairs,  the relevant contributions to the heat capacity
are due to electrons and collective modes. The collective mode contribution to the heat capacity can be obtained from the
dispersion relations we calculated in the preceding section. At low temperature when the typical momentum $k \ll k_{TF}$, the
dispersion relations are approximately linear with $\omega \simeq v'_\pm k$, and the associated heat capacity for each of the
collective modes is given by
\bea
C^\pm_V&=& \frac{2\pi^2}{15}~\left(\frac{k_BT}{\hbar v'_\pm}\right)^3~k_B \,, \nonumber \\
&\simeq &1.5 \times 10^{13}~\left(\frac{0.1 ~c}{v'_\pm}\right)^3~T^3_8~\frac{\rm ergs}{\rm cm^3~K}\,.
\eea
In contrast the electron contribution in the core
\bea
C^e_V&=& \pi^2~n_e~\frac{k_B^2T}{\mu_e}\,, \nonumber \\
&\simeq& 5.2 \times 10^{17}~\left(\frac{\mu_e}{100~ {\rm MeV}}\right)^2~T_8  ~\frac{\rm ergs}{\rm cm^3~K}\,,
\eea
is significantly larger. The contribution from the collective modes is small because of the higher power of the temperature and the relatively high velocities of the modes,  $\gtrsim 0.1 c$.

The coefficient of thermal conduction for a single species of bosonic mode with a dispersion relation $\omega_k$ is given by
standard kinetic theory
\be
\kappa=\frac{\hbar}{3T}\sum_{\bf k}\omega_k^2 \left(\frac{\partial \omega_k}{\partial k}\right)^2\left(-\frac{\partial
n_k}{\partial \omega_k}\right)\tau_k,
\label{kappa}
\ee
where $n_k=1/(\rme^{\hbar \omega_k/k_B T}-1)$ is the Bose distribution function, $\tau_k=1/(2\omega_k'')$ is the relaxation
time of the mode and $k_B$ is the Boltzmann constant.
For wave numbers small compared with the electron screening wave number, the velocity of the modes may be taken to be constant
and, on performing the sum in Eq.\ (\ref{kappa}), one finds the simple result
\bea
\nonumber
\kappa &=&\frac{\zeta(3)}{2\pi^2}\frac{(k_B T/\hbar)^2}{|v''|}k_B\\
\label{kappa_n}
&=&                       4.807\times 10^{10}  \frac{ T_{8}^2}{|v''|/c}  {\rm erg/ (cm\, sec\, K)}.
\eea
The thermal conductivity does not depend on the real part of the
velocity of the mode, and consequently the modes with the smallest $v''$  contribute most to heat transport.

It is interesting to compare thermal conductivity of bosonic excitations in superfluid neutrons given in Eq. (\ref{kappa_n}) with other contributions to the thermal conductivity.
Thermal conductivity of fermionic excitations in superfluid neutrons and the electron contributions were calculated in \cite{ShterninYakovlev2007}, and found to be of the order of $10^{24}$ erg/(cm sec K), which shows that the bosonic excitations provide a negligible contribution to the thermal conductivity.

The frequency spectrum of the longitudinal modes  that can be excited in the core is sensitive to the real part of the velocity in the long-wavelength limit and will in general differ from the sound velocity defined by $\sqrt{\partial P/\partial \rho}$, where
$P$ is total pressure and $\rho c^2$ is the total energy density. In the parlance of stellar oscillations these modes are referred to as pressure modes, or simply p-modes.  The existence of two acoustic modes in the long-wavelength limit is
a unique feature of the superfluid-superconducting multi-component system that we have studied here, and it
differs qualitatively from what is expected in a normal fluid where only one mode exists, with velocity $c_s$.
Further, the difference in mode velocities is small and this may lead to unique neutron star seismology.

Perturbations either during accretion, thermonuclear x-ray bursts and superbursts, and neutron stars mergers may excite modes
in the core but it is unclear if neutron star seismology can be probed observationally. One model of quasi-periodic
oscillations observed in the tails of magnetar giant flares invokes a  coupling between internal modes of the neutron star crust and the
global magnetic field to provide a mechanism by which surface emission can be modulated at characteristic frequencies \cite{QPO}. It
would be interesting to explore if a similar coupling between the magnetic field and core p-modes can lead to observable
effects.

\section*{Acknowledgments}
D.N.K. gratefully acknowledges the hospitality of the Niels Bohr Institute, Copenhagen, and of the Ioffe Physical-Technical Institute, Saint Petersburg. This work was supported in part by the Russian Fund for Basic Research Grant 31 16-32-60023/15, the European Research Council Grant No.\ 307986 STRONGINT, the US Department of Energy Grant No.\ DE-FG02-00ER41132, and by NewCompStar, COST Action MP1304.

\begin{appendix}

\section{Equivalence of two approaches to superfluid dynamics}
In the traditional approach to describing superfluidity \cite{LLFluidMech} one works in terms of the so-called superfluid
velocity, ${\bf v}_s={\bm\nabla} \phi/m$, which is really a momentum, and the velocity of the normal component.  This has
previously been applied to the superfluid neutrons in the inner crust of neutron stars \cite{CJPChamelReddy,
KobyakovPethick2013,ChamelPageReddy}.
In the present context, when both neutrons and protons are superfluid, it is natural to treat them symmetrically, as we have
done in the text.  However, as we shall demonstrate, the equations of motion are equivalent to those in the traditional
formalism.  For simplicity we consider the case of small perturbations about a uniform state with both components at rest.
We express  the equations of motion (\ref{Eulp1}) and (\ref{Euln1}) in terms of the average velocity of the protons, which is
the proton current density divided by the proton number density:
\be
\bar{\mathbf v}_p\equiv \frac{n_{pp}}{n_p}\frac{\mathbf{p}_p}{m} +\frac{n_{pn}}{n_p}\frac{\mathbf{p}_n}{m}.
\label{tildevp}
\ee
The continuity equations (\ref{contp}) and (\ref{contn}) for nucleons therefore become
\be
 \partial_t n_p+{\bm\nabla}\cdot (n_p \bar{\mathbf{v}}_p)=0,
 \label{contpApp}
\ee
and
\be
 \partial_t n_n+{\bm\nabla}\cdot\left[ \left(n_{nn}-\frac{n_{np}^2}{n_{pp}}\right) \frac{\mathbf{p}_n}m+\frac{n_pn
 _{np}}{n_{pp}}\bar{\mathbf{v}}_p  \right]=0.
\label{contnApp}
\ee
Equations (\ref{contpApp}) and (\ref{contnApp}) have the form to be expected for a two-fluid model, even though in the
present case the two components are both superfluid.  The neutron current density is given by
\be
{\bf j}_n=\left(n_{nn}-\frac{n_{np}^2}{n_{pp}}\right) \frac{\mathbf{p}_n}{m}+\frac{n_pn _{np}}{n_{pp}}\bar{\mathbf{v}}_p ,
\ee
and therefore the superfluid neutron density in the notation of Refs. \cite{CJPChamelReddy,KobyakovPethick2013} is given by
\be
n_n^s=n_{nn}-\frac{n_{np}^2}{n_{pp}},
\ee
while the density of neutrons entrained with the protons is given by
\be
n_n^n=\frac{n_pn _{np}}{n_{pp}}.
\label{nnn}
\ee
In the traditional approach, the superscript $s$ refers to the  {\underline s}uperfluid component and $n$ to the  {\underline
n}ormal component.  In the present case, both components are superfluid, and the superscript $n$ refers to the component
whose dynamics is described in terms of the average velocity of the component, rather than the momentum per particle in the
condensate.
The equation of motion for ${\mathbf p}_n$, Eq.\ (\ref{Euln1}) has the usual form as in the two-fluid model.  
We turn now to the equation of motion for $\bar{\mathbf v}_p$.  From Eqs.  (\ref{Eulp1}) and (\ref{Euln1}) it
follows that
\bea
\partial_t\bar{\mathbf v}_p=\frac{n_{pp}}{mn_p}\partial_t\mathbf{p}_p +\frac{n_{pn}}{mn_p}\partial_t\mathbf{p}_n\\
= -\frac{n_{pp}}{mn_p}{\bm\nabla}\left(\mu^{\rm nuc}_p +e \Phi  \right) -\frac{n_{pn}}{mn_p}{\bm\nabla}\mu^{\rm nuc}_n.
\label{dvtildep}
\eea

 The total normal mass density is given by
 \begin{eqnarray}
\rho^n=m(n_p +n_n^n).
\end{eqnarray}
From Eq.~(\ref{nnn}) one thus sees that
\begin{eqnarray}
\rho^n=  m n_p\left(1+\frac{n^{np}}{n^{pp}}\right) =m\frac{n_p^2}{n^{pp}},
\end{eqnarray}
where the latter result is a consequence of the condition (\ref{n_pp}) for Galilean invariance.
Thus Eq.~(\ref{dvtildep}) may be written in the form
\be
\rho^n\partial_t\bar{\mathbf v}_p
= -n_p{\bm\nabla}\left(\mu^{\rm nuc}_p +e \Phi  \right) -n_n^n{\bm\nabla}\mu^{\rm nuc}_n.
\label{dvtildep2}
\ee
This result is consistent with \cite{KobyakovPethick2013}, Eq. 25, for the inner crust if one there neglects the shear
rigidity of the solid.  
Equation (\ref{dvtildep2}) represents a generalization of earlier results \cite{CJPChamelReddy,ChamelPageReddy,KobyakovPethick2013} in that charge neutrality
is not enforced locally.
\end{appendix}

\end{document}